# Numerical Test and Analysis of the 2$^{\text{nd}}$ Law of Black Hole Thermodynamics with Gravitational-Wave Data from Binary Black Hole Merger Events

Joan Sonnenberg

joan.sonnenberg2718@gmail.com

January 12, 2022

**ABSTRACT**

Gravitational-wave detections of black hole mergers in binary systems offer an excellent opportunity to test the 2$^{\text{nd}}$ law of black hole thermodynamics. In this paper, we review how the entropy of any astrophysical black hole is calculated and we use LIGO and VIRGO's mass and spin data measurements from black hole merger events detected over the past years to perform entropy calculations and numerically test the generalized 2$^{\text{nd}}$ law of thermodynamics. Besides, we analyze the mathematical correlation between the black hole merger event's initial parameters to prove and conclude that the theorem will always hold.

## I. INTRODUCTION

Gravitational-wave (GW) events detected by LIGO and VIRGO have provided us real evidence of black hole mergers [Abbott 2016a]. Whenever the gravitational attraction between the two black holes that constitute a binary system is strong enough, the two massive objects culminate in a collision that originates a new more massive black hole. Such events are called black hole mergers and generate gravitational waves which we can detect since 2015 [Abbott 2016b]. Since gravitational waves happen to be perfect footprints in space-time of black hole merger events, we can obtain plenty of information on the latter. Among others, gravitational waves provide us information about the mass and angular momentum of the initial black holes that constituted the binary system as well as of the new black hole remnant of the merger.

The purpose of this article is to test and analyze the validity of a well-known theorem; the 2$^{\text{nd}}$ Law of Black Hole Thermodynamics. As we will see in section II, this theorem demands that the entropy sum of the two primary black holes in a binary system must always be smaller or equal than the entropy of the remnant black hole originated by the merger of the primary ones. In section III we will show how the entropy of all astrophysical black holes is completely parameterized by their mass and angular momentum. Accordingly, in this paper we collect black hole mass and angular momentum data measured from all the black hole merger events that have been detected since 2015 to calculate entropy values and perform a numerical test of the validity of the mentioned theorem.

## II. THE AREA-ENTROPY THEOREM

The 2$^{\text{nd}}$ Law of Thermodynamics (1) states that the total entropy $S$ of any isolated system can never decrease in any classical physical process and therefore will always increase or remain constant.

$$\Delta S \geq 0 \tag{1}$$

This law, in the particular case of black holes, happens to be not informative. If any object with entropy were to be swallowed by a black hole, the object's entropy would become invisible to an exterior observer (we cannot measure anything that comes from the inside of a black hole's event horizon) and from its point of view the 2$^{\text{nd}}$ law would have been violated, since the entropy of the black hole's outer region would have decreased. Combining the entropy of a black hole and the ordinary entropy of matter and radiation fields in the black hole's exterior region provides us a more useful law; the generalized 2$^{\text{nd}}$ Law of Thermodynamics (GSL) [Bekenstein 1972], [Hod 2015]. The GSL (2) states that the sum of the entropy of a black hole and the ordinary entropy $S_{\text{o}}$ outside the massive object never decreases and typically increases as a consequence of the generic transformations of the black hole.

$$\Delta S_{\text{o}} + \Delta S_{\text{BH}} \geq 0 \tag{2}$$

If matter containing entropy falls into a black hole, this theorem postulates that the increase of the black hole's entropy more than compensates for the disappearance of ordinary entropy outside the black hole for an external observer [Bekenstein 1973a].

Applied to binary black hole system merger events the generalized 2$^{\text{nd}}$ law describes that the sum of the individual entropies $S_1$ and $S_2$ of two black holes in a binary system must be equal or smaller than the entropy of the final black hole $S_{\text{f}}$ originated from the merger together with the entropy of the gravitational wave $S_{\text{GW}}$ produced during the collision of the two massive objects. As follows:

$$S_{\text{f}} + S_{\text{GW}} \geq S_1 + S_2 \tag{3}$$

Theoretically, gravitational waves cannot have entropy, and if

|  | Non-Rotating ($J=0$) | Rotating ($J>0$) |
|---|---|---|
| Uncharged ($Q=0$) | Schwarzschild | Kerr |
| Charged ($Q>0$) | Reissner-Nordström | Kerr-Newman |

**Table 1.** The four theoretical types of black holes that exist according to Einstein's field equations distinguished according to whether or not they rotate ($J$) and whether or not they are charged ($Q$).

they were to have, its value would be negligible [Unnikrishnan 2015]. Therefore:

$$S_\text{f} \geq S_1 + S_2 \tag{4}$$

Note that $S_\text{GW} = 0$, if any, makes the inequality more valid. Consequently, in order to perform a test of the GSL we need to calculate the entropy of the three black holes involved in a binary system merger event (BH$_1 \to S_1$, BH$_2 \to S_2$ and BH$_\text{f} \to S_\text{f}$).

To do so, we resort to the area-entropy theorem [Hawking 1976], [Bekenstein 1972] and [Bekenstein 1973b], which establishes the relationship between a black hole's event horizon surface area $A$ and its entropy $S$ (in natural units):

$$S_\text{BH} = \frac{1}{4}A \tag{5}$$

From equation (5) we observe that $S \propto A$ which means that a black hole's event horizon surface area is directly related to its corresponding entropy and calculating $A$ will allow us to calculate $S_\text{BH}$ and test the theorem. Note that $S \propto A$ implies that proving

$$A_\text{f} \geq A_1 + A_2 \tag{6}$$

is the same as proving equation (4).

**III. ENTROPY OF KERR BLACK HOLES**

According to the no-hair theorem [Johannsen and Psaltis 2011], all theoretical black holes can be completely described by only three externally observable classical parameters: mass $m$, electric charge $Q$ and angular momentum $J$. Since all black holes must have mass but do not necessarily have to have electric charge nor angular momentum, there are four different types of black holes as shown in Table 1.

Each one of these four theoretically possible black holes is mathematically described by a different metric. A metric is an exact solution to Einstein's field equations that describes how space-time behaves around massive objects (such as black holes). Thus, there are 4 different metrics (equations) that provide a mathematical model of a black hole; each one of which leads to the description of a black hole with different characteristics (as shown in Table 1).

Astrophysical charged black holes in the universe happen to be rapidly neutralized [Gong 2019] so we expect their charge $Q$ to be 0 or negligible. Consequently, we can discard the possibility of astrophysical black holes being of the Reissner-Nordström or Kerr-Newman kind. Moreover, Schwarzschild black holes are simply a particular case of Kerr black holes where there is no angular momentum ($J=0$). Therefore we can say that the Kerr metric provides an exact mathematical description of every black hole we observe in the universe [Kerr 2008].

The Kerr metric only describes black holes with event horizons when its parameter $\Delta = 0$ [Teukolsky 2015], where $\Delta$ is given as:

$$\Delta = \alpha^2 + r^2 - r_s r = 0 \tag{7}$$

where $\alpha = \frac{J}{mc}$, $r$ is the black hole's event horizon radius and $r_s = \frac{2Gm}{c^2}$ (Schwarzschild's radius).

Gravitational waves provide us information about the individual masses and angular momenta of the two primary black holes right before their merger and of the final black hole immediately after the merger. The data measured by LIGO and VIRGO corresponds to these two specific moments. As seen previously, in order for these black holes to have an event horizon we must let $\Delta = 0$. Continuing, from $\alpha^2 + r^2 - r_s r = 0$, we obtain a quadratic equation which we can solve to find the event horizon radius of a Kerr black hole:

$$r_\pm = \frac{r_s \pm \sqrt{r_s^2 - 4\alpha^2}}{2} \tag{8}$$

or, substituting:

$$r_\pm = \frac{Gm}{c^2} \pm \sqrt{\left(\frac{Gm}{c^2}\right)^2 - \left(\frac{J}{mc}\right)^2} \tag{9}$$

where $J$ is a black hole's angular momentum, $m$ its mass, $c$ the speed of light and $G$ the gravitational constant.

While $r_-$ is the solution for the inner horizon of a Kerr black hole, $r_+$ is the solution for the outer horizon and the one of interest for this test. Note that in the limit $J \to 0$ (non-rotating), the radius of the Kerr black hole becomes $r = \frac{2Gm}{c^2}$, the Schwarzschild radius.

Once determined the radius of the event horizon of a Kerr black hole, its event horizon surface area is given by the following equation [Pickett 2000]:

$$A = 4\pi \left[ r^2 + \left(\frac{J}{mc}\right)^2 \right] \tag{10}$$

As seen in section II, the surface area of the event horizon of a black hole is directly proportional to its entropy ($S \propto A$). Therefore, we use equation (5) to calculate the entropy of black holes [Hawking 1976]. In its appropriate units ($J/K$):

$$S_\text{BH} = \frac{Akc^3}{4\hbar G} \tag{11}$$

where we introduce $k$ as Boltzman's constant and $\hbar$ as Planck's reduced constant. This equation describes the total amount of entropy that must be assigned to a black hole. Moreover, we obtain that the entropy of a black hole depends entirely on its mass and angular momentum ($m$ and $J \to r \to A \to S$). Gravitational wave detectors can provide us with black-hole data on the mentioned magnitudes. We measure solar mass units ($M_\odot$) and information on the angular momentum of black holes is given via the dimensionless spin parameter $a = \frac{cJ}{Gm^2}$. A Kerr black hole's maximum angular momentum is $a=1$ (or $a=-1$, if spinning in the completely opposite direction). $a=0$ is a non-spinning black hole (Schwarzschild). Substituting in equation (5) with the previously derived formulas for a black hole's event horizon radius $r$ and area $A$ we conclude that the entropy of any astrophysical black hole in the universe is given by the equation:

$$S_\text{BH} = \frac{2\pi k G m^2 (1 + \sqrt{1-a^2})}{c\hbar} \tag{12}$$

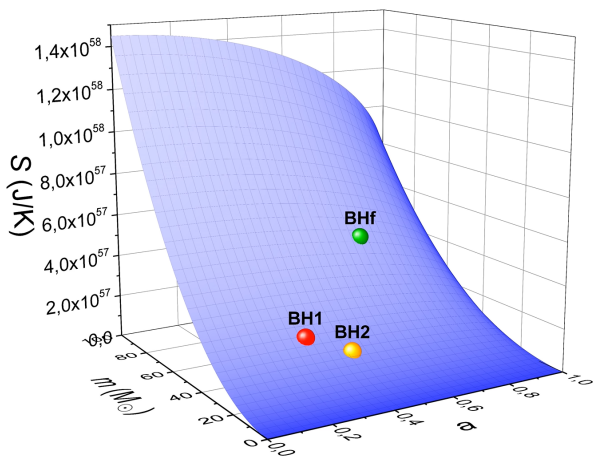

**Figure 1.** The entropy of all astrophysical black holes in our universe as a function of their mass $m$ (in Solar mass units $M_\odot$) and dimensionless spin parameter $a$. Although the spin parameter can take values from -1 to 1 we only show for positive values of $a$, since, until today, all spin measurements of black holes have been within this range. In red, yellow and green we highlight the particular calculated entropies of the three black holes involved in the merger event GW150914.

In Figure 1 we show how entropy $S$ behaves as a function of the mass $m$ and spin $a$ of a black hole. As an example, we have highlighted three specific dots corresponding to the entropy of the three black holes involved in the black hole merger event GW150914. In red and yellow we show the primary black holes $BH_1$ and $BH_2$ while in green the remnant black hole $BH_f$.

## IV. GW DATA AND TEST

In this section, we test the generalized 2$^{nd}$ law of thermodynamics using LIGO and VIRGO mass and spin parameter data measurements from their first three observational runs from 2015, 2017 and 2019 [Abbott 2018a], [Abbott 2019] and [Abbott 2021]. During these, several gravitational-wave detections were made and proved to be coming from black hole-merger events from binary systems.

Both the mass and spin parameters characterizing a black hole leave an imprint on the gravitational-wave signal during a coalescence. Therefore, LIGO and VIRGO have been able to obtain the mass and spin values of the primary black holes from the binary system and of the final black hole resulting of the merger. With these data, we can calculate the entropy of the individual three black holes involved in a merger event using equation (12) and test the generalized second law of thermodynamics by verifying that inequality (4) holds.

To date, there are many confirmed events of black hole mergers, 12 of which we have sufficient data to test and analyze the GSL. The merger events detected during LIGO and VIRGO's first and second observing runs ($O_1$ and $O_2$) are GW150914, the first-ever gravitational wave detection, GW151012, GW151226, GW170104, GW170608, GW170729, 170809, GW170814, GW170818 and GW170823. Additionally, from their third observing run ($O_3$) we have GW190412 and GW190521, the most massive black hole merger until now. Each black hole's mass $m$ and spin $a$ data from the merger events used for this test are shown in Table 3. Performing the entropy calculations for each black hole using equation (12) and the data mentioned and collected we obtain Figure 2. We compare the calculated sum of entropies of the two primary black holes $S_1 + S_2$ (plotted in black in the Figure 2) to the entropy of the final black hole (plotted in red) and conclude that the theorem is true since the inequality $S_f \geq S_1 + S_2$ always holds within the error.

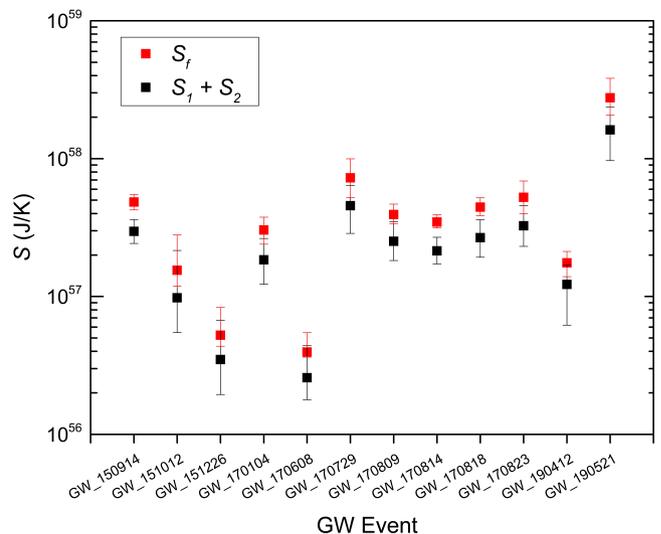

**Figure 2.** Numerical proof of the GSL that confirms its validity. For the 12 LIGO-VIRGO confirmed black hole binary system merger events, we plot in red the entropy of the remnant black holes $S_f$ compared to the entropy sum (in black) of the primary black holes ($S_1 + S_2$) of each merger event.

| GW 150914 | $m$ | $a$ | $S$ |
| --- | --- | --- | --- |
| $BH_1$ | $35.8^{+5.3}_{-3.9}$ | $0.32^{+0.49}_{-0.29}$ | $\left(1.81^{+0.557}_{-0.404}\right) \times 10^{57}$ |
| $BH_2$ | $29.1^{+3.8}_{-4.3}$ | $0.44^{+0.50}_{-0.40}$ | $\left(1.16^{+0.339}_{-0.365}\right) \times 10^{57}$ |
| $BH_f$ | $62.0^{+4.1}_{-3.7}$ | $0.67^{+0.05}_{-0.07}$ | $\left(4.85^{+0.654}_{-0.605}\right) \times 10^{57}$ |

**Table 2.** Mass $m$ ($M_\odot$) and spin $a$ data values of the three black holes ($BH_1$, $BH_2$ and $BH_f$) involved in the merger event GW150914 [Abbott 2016b]. Additionally, the calculated entropy $S$ corresponding to each one of them (calculated using equation 12).

For instance, we will show for the merger event GW150914 [Abbott 2016b]. Table 2 shows the data parameters of the three black holes involved in the merger. Calculating the entropy for each one of them we obtain (in ($J/K$) units):

$$(4.85 \pm 0.956) \times 10^{57} > (2.97 \pm 0.598) \times 10^{57} \qquad (13)$$

The inequality holds hence the merger event GW150914 obeys the generalized 2$^{nd}$ law of thermodynamics. Nevertheless, it is quite noticeable that source parameter errors are quite significant and thus, must be taken into account. We drag the mass and spin error margins to present the error bounds for the calculated entropy of each black hole. We use the error drag formula according to which if $V$ is a function of the variables $x$ and $y$ with their respective error bounds $\Delta x$ and $\Delta y$:

$$V = V(x, y, ...) \qquad (14)$$

| GW Event | $m_1$ | $m_2$ | $m_{\rm f}$ | $a_1$ | $a_2$ | $a_{\rm f}$ | Ref. |
|---|---|---|---|---|---|---|---|
| GW 150914 | $35.8^{+5.3}_{-3.9}$ | $29.1^{+3.8}_{-4.3}$ | $62.0^{+4.1}_{-3.7}$ | $0.32^{+0.49}_{-0.29}$ | $0.44^{+0.50}_{-0.40}$ | $0.67^{+0.05}_{-0.07}$ | [Abbott 2016b] |
| GW 151012 | $23^{+18}_{-6}$ | $13^{+4}_{-5}$ | $35^{+14}_{-4}$ | $0.31^{+0.48}_{-0.28}$ | $0.45^{+0.48}_{-0.41}$ | $0.66^{+0.09}_{-0.10}$ | [Abbott 2018a] |
| GW 151226 | $14.2^{+8.3}_{-3.7}$ | $7.5^{+2.3}_{-2.3}$ | $20.8^{+6.1}_{-1.7}$ | $0.49^{+0.37}_{-0.42}$ | $0.52^{+0.43}_{-0.47}$ | $0.74^{+0.06}_{-0.06}$ | [Abbott 2018a] |
| GW 170104 | $31.2^{+8.4}_{-6.0}$ | $19.4^{+5.3}_{-5.9}$ | $48.7^{+5.7}_{-4.6}$ | $0.45^{+0.46}_{-0.40}$ | $0.47^{+0.46}_{-0.43}$ | $0.64^{+0.09}_{-0.20}$ | [Abbott 2018b] |
| GW 170608 | $11.3^{+5.6}_{-2.0}$ | $7.5^{+1.5}_{-2.2}$ | $17.8^{+3.4}_{-0.7}$ | $0.32^{+0.47}_{-0.29}$ | $0.43^{+0.49}_{-0.39}$ | $0.69^{+0.04}_{-0.04}$ | [Abbott 2019], [Soumi 2019] |
| GW 170729 | $49.5^{+12.1}_{-10.2}$ | $32.2^{+9.9}_{-9.1}$ | $79.5^{+14.7}_{-10.2}$ | $0.60^{+0.34}_{-0.51}$ | $0.57^{+0.38}_{-0.50}$ | $0.81^{+0.07}_{-0.13}$ | [Abbott 2019], [Soumi 2019] |
| GW 170809 | $35.0^{+9.1}_{-5.9}$ | $23.9^{+5.0}_{-5.3}$ | $56.3^{+5.2}_{-3.8}$ | $0.34^{+0.53}_{-0.31}$ | $0.40^{+0.51}_{-0.37}$ | $0.70^{+0.08}_{-0.09}$ | [Abbott 2019], [Soumi 2019] |
| GW 170814 | $30.4^{+5.7}_{-2.7}$ | $25.8^{+2.6}_{-4.0}$ | $53.2^{+3.2}_{-2.4}$ | $0.53^{+0.42}_{-0.48}$ | $0.46^{+0.47}_{-0.42}$ | $0.72^{+0.07}_{-0.05}$ | [Abbott 2019], [Soumi 2019] |
| GW 170818 | $36.1^{+8.5}_{-5.3}$ | $26.5^{+4.7}_{-6.0}$ | $59.4^{+4.9}_{-3.8}$ | $0.56^{+0.38}_{-0.50}$ | $0.50^{+0.44}_{-0.45}$ | $0.67^{+0.07}_{-0.08}$ | [Abbott 2019], [Soumi 2019] |
| GW 170823 | $39.2^{+10.9}_{-6.6}$ | $28.9^{+6.3}_{-7.2}$ | $65.4^{+10.1}_{-7.4}$ | $0.44^{+0.46}_{-0.40}$ | $0.45^{+0.48}_{-0.41}$ | $0.72^{+0.09}_{-0.12}$ | [Abbott 2019], [Soumi 2019] |
| GW 190412 | $29.18^{+5.84}_{-7.73}$ | $8.56^{+2.74}_{-1.10}$ | $37.3^{+3.9}_{-3.8}$ | $0.56^{+0.19}_{-0.21}$ | $0.56^{+0.39}_{-0.50}$ | $0.67^{+0.05}_{-0.06}$ | [Abbott 2021], [Gerosa 2020] |
| GW 190521 | $91^{+29}_{-16}$ | $67^{+18}_{-21}$ | $150^{+29}_{-18}$ | $0.71^{+0.26}_{-0.59}$ | $0.60^{+0.36}_{-0.53}$ | $0.73^{+0.11}_{-0.14}$ | [Abbott 2020] |

**Table 3.** Data collected from LIGO and VIRGO's observing runs $O_1$, $O_2$ and $O_3$ we used to perform the test of the area-entropy theorem (GSL). Columns show source primary mass $m_1$, secondary mass $m_2$, final mass $m_{\rm f}$, dimensionless primary spin parameter $a_1$, dimensionless secondary spin parameter $a_2$ and final spin $a_{\rm f}$ of each merger event detected. Mass parameter values in solar mass units ($M_\odot$). Median and 90% symmetric interval values of the source parameters taken from the catalogs of gravitational wave detections cited in column Ref.

the error bounds of $V$ are given by:

$$\Delta V = \sqrt{\left(\frac{dV}{dx}\right)^2 \cdot \Delta x^2 + \left(\frac{dV}{dy}\right)^2 \cdot \Delta y^2 + \cdots} \quad (15)$$

Even though we have previously concluded that Figure 2 proves that $S_{\rm f} \geq S_1 + S_2$ works for all tested merger events, some of them might not always obey the GSL when considering the error bounds (notice that error bars overlap). We can completely be certain of the validity of the theorem for the merger events GW150914, GW170814 and GW170818 since the inequality is always satisfied (even considering the worst-case scenario error bounds). Meanwhile, this does not happen with the rest of the merger events in which the entropy error bounds are much larger. We conclude that the generalized 2$^{\rm nd}$ law of thermodynamics is empirically proven since all mergers obey the inequality (within the error). However, future black hole merger events measured with more precision (measurements obtaining smaller error bounds for $m$ and $a$) thanks to advances in technology and measurement techniques of gravitational waves will allow us to perform more solid tests of the GSL's validity if not prove a violation of said theorem, which would of course be highly significant and paradigm-changing.

## V. RESULTS AND DISCUSSION

In this section, we analyze the mass ($m$) and spin ($a$) data values from Table 3 of the black holes involved in all the BH merger events to encounter if there is any kind of mathematical relationship between these parameters that affects entropy ($S$) and makes the GSL theorem hold in all the merge events tested.

In Figure 3a we represent the loss of total mass in black hole binary systems during their merger events. We plot the amount of mass lost in the system as $\Delta m = m_{\rm f} - (m_1 + m_2)$ with respect to the initial mass of the system ($m_1 + m_2$). During merger events, a certain quantity of the mass of the primary black holes is lost in the form of energy ($E = mc^2$) which generates the gravitational waves LIGO and VIRGO have detected. Consequently, and as we can observe, $\Delta m$ will always be negative meaning that the remnant black hole of the merger will never be more massive than the mass sum of the two primary black holes ($m_1 + m_2 > m_{\rm f}$). Note that in the graph, the more massive the system, the greater the loss of mass during the merger event. We observe a linear correlation that fits the data values following the equation:

$$\Delta m = 0.29 - 0.05(m_1 + m_2) \quad (16)$$

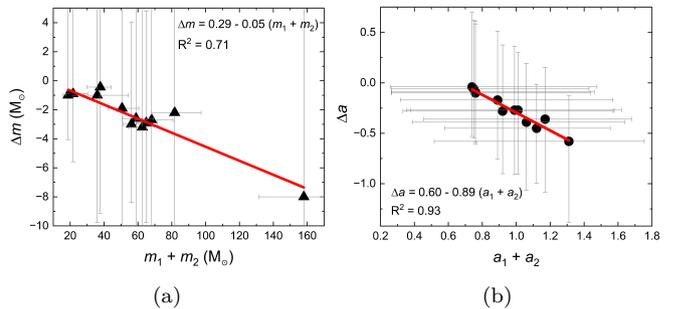

**Figure 3.** In Figure (a) we show the correlation we have observed between the primary mass of a black hole binary system ($m_1 + m_2$) and the mass of the remnant black hole resulting from the merger event ($M_{\rm f}$). We present this correlation in the form of $\Delta m$ which tells us the mass lost in the system in the form of energy during the merger event. In Figure (b) we show the same correlation but this time with the spin parameter. The plot represents the total amount of spin in the system before the merger event ($a_1 + a_2$) and the spin of the remnant black hole after the merger ($a_{\rm f}$). We also present the correlation in the form of $\Delta a$ to describe the amount of spin lost during the merger event.

Therefore, knowing the total mass of a binary black hole system is enough to predict the mass of the remnant black hole after the merger occurs. Rearranging:

$$m_{\rm f} = 0.95 m_1 + 0.95 m_2 + 0.29 \quad (17)$$

Something similar happens with the spin parameter. In Figure 3b we plot the loss of the total amount of spin in black-hole binary systems as $\Delta a = a_{\rm f} - (a_1 + a_2)$ with respect to the spin sum of the two primary black holes in the system ($a_1 + a_2$). Although the remnant black hole's spin is larger than the spin of the individual primary black holes (such as happens with the mass), it is not larger than the sum of them

$(a_1+a_2)$; meaning $a_1+a_2 > a_\text{f}$. Hence, $\Delta a$ will always be negative. As we can observe, the amount of "lost spin" increases as the spin sum of the primary black holes does $(a_1 + a_2)$ following a linear correlation that fits the data values following the equation:

$$\Delta a = 0.60 - 0.89(a_1 + a_2) \qquad (18)$$

Such as happens with the mass, knowing the spin parameters of the two primary black holes in a system is enough to predict the remnant's spin value after the merger event. As follows:

$$a_\text{f} = 0.11 a_1 + 0.11 a_2 + 0.60 \qquad (19)$$

We conclude that we do not need mass and spin data of the remnant black holes of the merger events since both parameters can be predicted knowing only the data of the primary black holes in their systems.

As a final result, in Figure 4 we show how the entropy of remnant black holes from merger events ($S_\text{f}$) is always larger than the sum of the individual entropies of the two primary black holes ($S_1 + S_2$). We plot the difference between the entropy of the remnant black hole and the entropy sum of the individual primary black holes as $\Delta S = S_\text{f} - (S_1 + S_2)$ with respect to the initial entropy in the binary system ($S_1 + S_2$). Thus, as long as $\Delta S$ (which tells the extra entropy that $S_\text{f}$ has compared to $S_1 + S_2$) remains positive, the generalized 2$^\text{nd}$ law of thermodynamics is valid. Note that, as the entropy sum of the primary black holes increases, so does $\Delta S$. Consequently, the difference between $S_\text{f}$ and $S_1 + S_2$ gets bigger implying more certainty of the theorem being true. From the plot of the calculated entropy values of the black holes in the 12 merger events we observe a linear regression relationship between $\Delta S$ and $S_1 + S_2$ that follows the equation:

$$\log \Delta S = -4.17 + 1.07 \log(S_1 + S_2) \qquad (20)$$

With an extremely convincing regression coefficient of $R^2 = 0.96$ (within the error) equation (20) predicts which will be the value of $\Delta S$ for every possible $S_1 + S_2$ we can imagine in the universe from any binary black hole merger event. Rearranging the equation we obtain:

$$S_\text{f} = \frac{1}{10^{4.17}}(S_1 + S_2)^{1.07} + (S_1 + S_2) \qquad (21)$$

Note that as $S_1 + S_2$ gets smaller, so does $S_\text{f}$, but the latter always remains larger than $S_1 + S_2$. Even when the entropy sum of the primary black holes tends to 0, the limit for $S_\text{f}$ is also 0:

$$\lim_{S_1+S_2 \to 0} \frac{1}{10^{4.17}}(S_1+S_2)^{1.07} + (S_1+S_2) = 0 \qquad (22)$$

meaning that even in this case the theorem is valid (since $\Delta S \geq 0$). Contrarily, as $S_1 + S_2$ increases, so does $\Delta S$, so the theorem always holds.

From this numerical analysis of the generalized 2$^\text{nd}$ law of thermodynamics we conclude that the theorem will always hold no matter what the initial parameters are of the black holes within any binary system. Therefore, we have proven it and we confirm its validity.

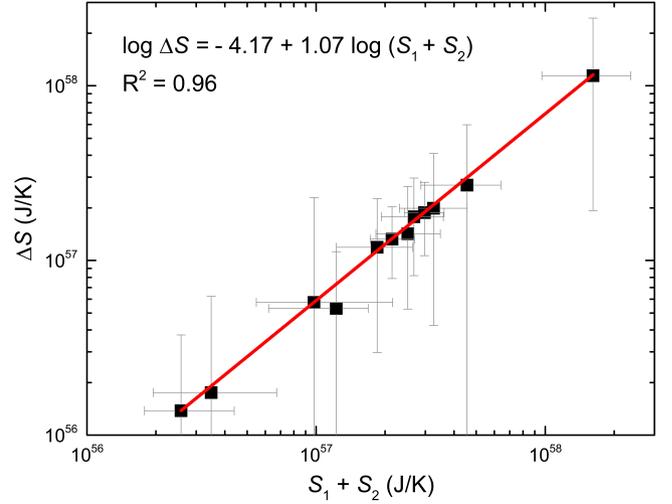

**Figure 4.** Difference between the entropy of the remnant black hole and the initial entropy in the system $\Delta S = S_\text{f} - (S_1 + S_2)$ with respect to the sum of the entropies of the primary black holes ($S_1 + S_2$). This plot proves that the GSL holds for the tested black hole merger events and that there exists a mathematical correlation between the entropies involved in a black hole merger event.

## Acknowledgements

I would like to thank Laia Casamiquela and Víctor Moreno for their extremely helpful advice and guidance during the writing of this paper. Besides, I would also like to mention and especially thank *Fundació Catalunya La Pedrera* for offering me the unique opportunity to participate in their *Youth and Science* program which has introduced me to scientific research and has led me to the writing of this paper.